\documentclass[aps,pra,twocolumn,longbibliography]{revtex4-1}

\usepackage{graphicx}
\usepackage{braket}
\usepackage{amsmath}
\usepackage{amssymb}
\usepackage{amsthm}

\begin{document}
\title{Practical Quantum Key Distribution with Geometrically Uniform States}
\author{K. S. Kravtsov$^{1,2}$}\email{kk@quantum.msu.ru}
\author{S. N. Molotkov$^{1,3,4,5}$}
\affiliation{
$^1$ Quantum Technology Centre of Moscow State University, Moscow, Russia\\
$^2$ A.M.Prokhorov General Physics Institute RAS, Moscow, Russia \\
$^3$Institute of Solid State Physics RAS, Chernogolovka, Moscow region, Russia  \\
$^4$Academy of Cryptography of Russian Federation, Moscow, Russia \\
$^5$Faculty of Computational Mathematics and Cybernetics, Moscow State University, Moscow, Russia \\
}

\date{\today}
\begin{abstract}
In this paper we propose a practical quantum key distribution protocol based on geometrically uniform states and a standard decoy state technique. The protocol extends the ideas used in SARG04 to the limit where the core quantum communication is secure against unambiguous state discrimination and provides some level of inherent resistance to photon number splitting attacks. An additional decoy state overlay ensures its conformity to the conventional security requirements. The protocol security is analyzed by an explicit construction of the most optimal unitary attack, which is known to reach the tight security bound in the case of BB84.
\end{abstract}
\maketitle

\section{Introduction}
Quantum cryptography is an actively developing technology that enables provably secure secret key exchange between remote parties. Roughly speaking, its security comes from the use of discrete elementary particles, usually photons, as information carriers. At the same time,  practical systems rely upon weak coherent pulses (WCPs) instead of true single photons mainly due to the present technological limitations. That creates a whole bunch of security issues compared with the perfectly secure original single-photon-based BB84 protocol~\cite{BB84}. The well-known threats are the photon-number-splitting (PNS) attack~\cite{HIG95} and the unambiguous state discrimination (USD) in the channel~\cite{BLM00}.

Construction of an ultimately secure WCP-based quantum key distribution (QKD) protocol is one of the central goals of  the area. However, up to the present day, there is no universal solution for practical WCP-based strategies. Currently, the approach based on the decoy state BB84 is widely considered to be the most appropriate for implementations.

It is well known that WCP-based implementation of the plain BB84 is insecure starting with quite moderate loss levels~\cite{BLM00}, while better alternatives were developed as early as 2004~\cite{SAR04}. The goal of the current paper is to extend the best solution from that time and combine it with the now conventional decoy state technique. The resulting QKD protocol is secure not only due to the employed decoy states, but is also immune to attacks due to its superior inner structure.

In the following text we will assume a phase encoding scheme, which better suits fiber based implementations. In this scheme each source pulse is split into two equal halves at different time slots and the relative optical phase between them encodes the information. We also assume that source WCPs have randomized optical phases. The paper is organized as follows: first we define a QKD protocol with geometrically uniform states (GUSs) and find a set of parameters convenient for practical implementations. Second, we discuss the structure of the transmitted quantum states and analyze its security against an optimal unitary attack. Finally, we apply the decoy state overlay and find the security bounds for the resulting QKD protocol. In the last sections we calculate key generation rates under typical conditions and discuss the obtained results.

\section{QKD protocol with GUS}
\begin{figure*}[tb]
\begin{center}
\includegraphics[width=0.7\textwidth]{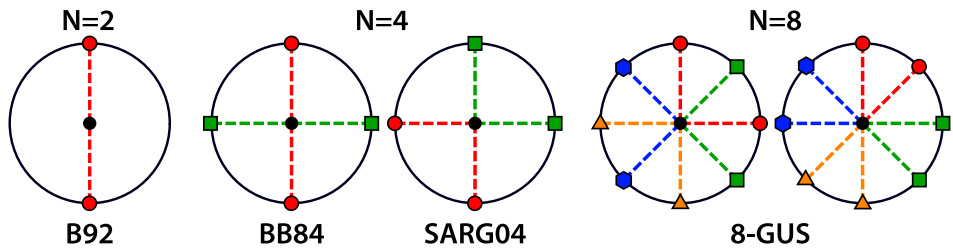}
\end{center}
\caption{
Information states and logical bases in different GUS protocols shown as phasor diagrams. The bases are displayed in different colors and marker shapes. 8-GUS protocols have two convenient choices of bases: with the angle between states of $\Delta\varphi=\pi/2$ or $\Delta\varphi=\pi/4$ as shown in the last two diagrams respectively.
}\label{fig_GUstates}
\end{figure*}

The protocol with GUS uses $N$ distinct geometrically uniform quantum states defined by a unitary transformation $U$, such as $U^N=I$. So $\ket{\psi_j} =  U^j \ket{\psi_0}$. In the case of phase encoding $\ket{\psi_j} = \ket{\alpha}_1 \otimes \ket{e^{i\frac{2\pi }{N}j}\alpha}_2,$ indices 1 and 2 belong to the two time slots, where the coherent states $\ket{\alpha}$ are located. It is clear that this general framework includes such protocols as B92~\cite{B92} where $N=2$, and BB84~\cite{BB84} and SARG04~\cite{SAR04}, where $N=4$.
As in BB84, the states are grouped by pairs into $N/2$ logical bases: each basis contains two states, with a certain angle between them, that correspond to the logical 0 and 1. $N$ is considered to be even. Examples of different GUS protocols with certain bases choices are shown in Fig.~\ref{fig_GUstates}.

The protocol proceeds as follows:
\begin{enumerate}
\item \emph{State generation and transmission.} Alice randomly selects a bit value of 0 or 1. She also randomly chooses one of the $N/2$ available bases.
Then the corresponding quantum state is generated and sent into the channel. All quantum states differ only by the relative phase between the two coherent states. This phase takes one of the $N$ possible values and is totally defined by the basis and the bit value. Although, it could be an arbitrary $N$ to $N$ mapping, we limit our analysis to the cases where there is a fixed phase shift $\Delta\varphi$ between 0 and 1 in the same basis, similar to the examples shown in Fig.~\ref{fig_GUstates}.
\item \emph{State measurement.} Bob randomly chooses one of the $N/2$ bases and
one of the two settings of the B92-like measurement in the chosen basis. For the explicit measurement procedure, see section~\ref{sec_proto_perf}. Then he performs the corresponding measurement.
\end{enumerate}

The steps 1 and 2 are repeated multiple times. Apparently, if Alice and Bob happen to use the same basis, Bob's detector will only click if his measurement setting matches the bit value sent by Alice. If bases differ, there still might be some correlation between the two, but, for simplicity of the analysis, we ignore it and discard those cases as unsuccessful.

\begin{enumerate}
	\setcounter{enumi}{2}
	\item \emph{Key sifting.} Bob tells Alice for which states his detector clicked and discloses the bases he used for his measurements. Alice, in reply, tells Bob for which states her basis matched the one Bob used. The bit values associated with the selected states thus constitute the {\it raw key}. All other data are discarded.
\end{enumerate}

After the whole procedure, Alice and Bob end up with the highly correlated raw keys, for which standard classical strategies of error correction and privacy amplification are used. The rest of the paper analyzes the accessible share of secret information contained in these correlated bit sequences.

As the phase $\theta$ of the carrier coherent state $\ket{\alpha}=\ket{e^{i\theta}|\alpha|}$ is completely random, the information states in the channel are mixed states with randomized phases, described by a density matrix
\begin{equation}
\rho=
\int_{0}^{2\pi} \frac{d \theta}{2\pi}
\ket{\psi}\bra{\psi}.
\end{equation}
The same density matrix is obtained~\cite{AGS04} with an incoherent mix of number states with Poissonian probability distribution:
\begin{equation}
\rho(\varphi)=
\sum_{k=0}^{\infty}
P_k(\mu)
|\psi_k(\varphi)\rangle\langle \psi_k(\varphi)|
,
\quad
P_k(\mu)= e^{-2\mu} \frac{(2\mu)^k}{k!},
\label{eq_fock_representation}
\end{equation}
\begin{equation}
|\psi_k(\varphi)\rangle=
\sqrt{\frac{k!}{2^k}}
\sum_{m=0}^k
\frac{e^{im\varphi}|m\rangle_1\otimes |k-m\rangle_2}{\sqrt{m!(k-m)!}},
\end{equation}
where $\mu = |\alpha|^2$ is the mean photon number in each of the two coherent states.
Due to the physical indistinguishability of the two interpretations, we are free to choose the second one as more convenient for further analysis.

\begin{figure}[tb]
\begin{center}
\includegraphics[width=\columnwidth]{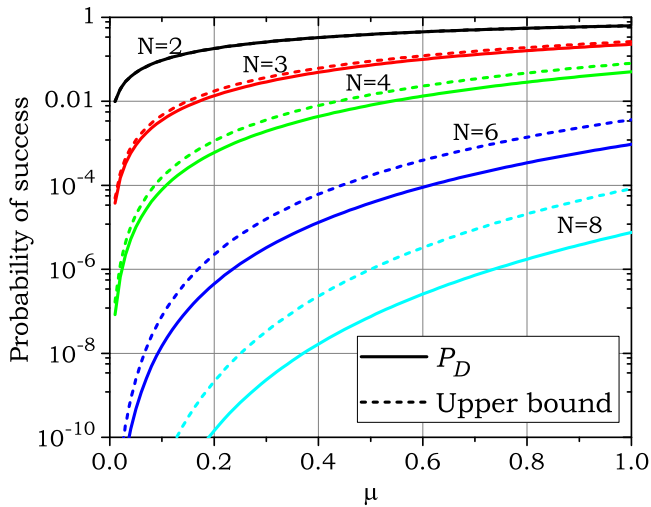}
\end{center}
\caption{
Probability of successful error-free discrimination of $N$ GUS as a function of the mean photon number $\mu$. The solid line shows the exact result, while the dashed line is the trivial upper bound given by~(\ref{eq_Pd-upper_bound}).
\label{fig_USD_probabilities}
}
\end{figure}

The proper choice of the alphabet size $N$ is closely related to the protection of the system against the USD attack. In general, the error-free discrimination between $N$ quantum states is possible when the states are linearly independent~\cite{C98}. If we consider the number state components of (\ref{eq_fock_representation}), USD of $N$ states is possible only for $k\ge N-1$~\cite{C01}. Therefore the probability of success $P_D$ for error-free measurement is necessarily smaller than the share of these components, i.e.
\begin{equation}P_D < \sum_{k=N-1}^\infty P_k(\mu).\label{eq_Pd-upper_bound}\end{equation}
More accurate values for $P_D$ may be found using the theory for the geometrically uniform coherent states~\cite{C98,CB98}. The corresponding probabilities calculated for common choices of $N$ as a function of the mean photon number $\mu$ are plotted in Fig.~\ref{fig_USD_probabilities}.

If $P_D$ is larger than Bob's probability of detection, Eve could in principle learn the whole key without producing errors by doing the USD and re-transmission of successfully measured states. For practical systems where the overall system efficiency is at least $10^{-6} \dots 10^{-5}$ it is safe to use $N=8$, while the common choice of $N=4$ cannot be considered secure. Presumably, the USD attack may be detected with the decoy state approach, however, to the best of our knowledge, there are no known bounds for USD attacks under the decoy state strategy. It has to be mentioned that an implementation of the USD attack does not require non-demolition measurements of the photon number as well as any quantum memory. Thus, it could be much more affordable for potential implementation than, for example, the PNS attack discussed later. Following this argument, we consider the case of $N=8$ to be of the most practical importance, while $N=16$ could be used for future implementations where much lower system efficiencies up to $10^{-12}$ could be safely tolerated.

The choice of the angle between states in a logical basis is also directly related to the system security. As discussed in~\cite{SAR04}, the use of non-orthogonal states makes the protocol much more secure, compared with the orthogonal case of e.g. BB84. The difference becomes apparent if we consider an adversary implementing the PNS attack, after which she can keep the exact copy of each delivered photon in her quantum memory. After the basis reconciliation, all logical bases become known and if the states are orthogonal it is possible to deterministically measure them and learn the whole key. If, on the contrary, the states are non-orthogonal, the adversary will only have limited information about the states due to their non-orthogonality. The probability of successful error-free discrimination between the two states in the logical basis is bounded by \begin{equation}p_k^\mathrm{USD} = 1 - \left|\braket{\psi_k(\varphi_0)|\psi_k(\varphi_1)}\right| = 1-\left[\cos \left( \frac{\Delta \varphi}{2} \right) \right]^k,\label{eq_p_usd}\end{equation}
where $\Delta \varphi=\varphi_1-\varphi_0$ is the angle between the logical zero and one, and $k$ is the number of available photons. For the case of the 8-GUS protocol, the convenient choice of the angle $\Delta \varphi$ is $\pi/2$ or $\pi/4$. The former gives higher key generation rates, while the latter can tolerate higher bit error ratios.

\section{Protocol performance under unitary attack}\label{sec_proto_perf}
So far we discussed the protocol behaviour under zero-error attacks, where Eve could learn the whole key without causing errors. That is, however, not enough for a quantitative analysis of the protocol security in a general case.

A unitary attack is a construction of an optimal unitary operator, which creates quantum entanglement between the information states in the channel and an auxiliary quantum state (Eve's ancilla). The modified information states propagate further down the line to Bob, while the ancillae are stored in Eve's quantum memory until bases are disclosed during the reconciliation stage. At this time Eve can perform a collective measurement over all her quantum memory to get the maximal amount of information about the key. The optimal strategy for Eve is to maximize this information given a fixed value of the observed quantum bit error ratio (QBER). It should be mentioned~\cite{FGG97} that {\em unitarity} of the attack does not lead to any loss of generality, as any physical non-unitary action may be described by a unitary evolution of a larger system. The same adversary strategy was used earlier to construct an explicit attack that reaches the tight security bound of $Q_c\approx 11\%$ for the BB84 protocol~\cite{M07}.

The engineering of an optimal eavesdropping operator is quite a complicated task, while calculation of a conservative upper bound may be done much more
easily. For the sake of this calculation we assume that Eve knows logical bases at the time of her interaction with the states in the channel, so her goal is to create a unitary operator for optimal discrimination between the two non-orthogonal states within the basis. It is known~\cite{FGG97,M07} that optimal strategy for Eve may be attained by a symmetric unitary attack, where roles of logical zeros and ones may be swapped. That also means that bit errors, observed by Bob follow properties of a {\em binary  symmetric channel}. Under that assumption we may write the following explicit form of this attack.

Denote $\ket{\Psi_k(0)} = \ket{\psi_k(\varphi_0)}$,  $\ket{\Psi_k(1)}=\ket{\psi_k(\varphi_1)}$, and $c_k = \big|\braket{\Psi_k(0)|\Psi_k(1)}\big|$ in the known logical basis. After the performed entangling operator, global quantum states are written as
\begin{equation}
\begin{split}
|\Phi_k(0)\rangle_{BE} ={} & U_{BE}( |\Psi_k(0)\rangle_B\otimes |E\rangle_E   )\\
={} & A_k |\Psi_k(0)\rangle_B\otimes |E_k(0)\rangle_E \\
& + B_k |\Psi_k(1)\rangle_B\otimes |E_k(1)\rangle_E,\\
|\Phi_k(1)\rangle_{BE}= {}&
U_{BE}( |\Psi_k(1)\rangle_B\otimes |E\rangle_E   )\\
={} & A_k |\Psi_k(1)\rangle_B\otimes |E_k(1)\rangle_E  \\
&+ B_k |\Psi_k(0)\rangle_B\otimes |E_k(0)\rangle_E,
\end{split}\label{eq_unitary_op}
\end{equation}
where  $U_{BE}$ is Eve's unitary action, $|E\rangle_E$ --- Eve's ancilla, and $|\Phi_k(0,1)\rangle_{BE}$ is the resulting entangled state of Eve and Bob; 
$|E_k(0,1)\rangle_E $ is the modified ancilla, and the parameters $A_k$ and  $B_k$ are defined by the choice of the unitary operator.
The goal of Eve is to achieve a maximal separation between $\ket{E_k(0)}$ and $\ket{E_k(1)}$, which is defined by their scalar product $c_k^\mathrm{E} = \big|{}_{E}\langle E_k(0) |E_k(1)\rangle_E\big|$.
When modified states arrive to Bob, we may consider them to be mixed states with traced out Eve's subsystem:
\begin{equation}
\rho_k(0,1)=
\mbox{Tr}_E  \Big[ |\Phi_k(0,1)\rangle_{BE}{}_{BE}\langle |\Phi_k(0,1) |    \Big].
\end{equation}

Next, we define Bob's measurement procedure to find the relation of $c_k^\mathrm{E}$ and the observed probability of error on Bob's side. Although, unperturbed states may be discriminated by an optimal POVM with the probability~(\ref{eq_p_usd}), practical systems use a suboptimal projective measurement approach. The phase of the receiving delay interferometer is adjusted to detect either logical zero or logical one, so the performed projective measurement is the following resolution of identity
\begin{equation}
I_k = |\Psi_k(j)\rangle_B{}_B\langle \Psi_k(j)| + |\Psi_k^\perp (j)\rangle_B{}_B\langle \Psi_k^\perp (j)|,
\end{equation}
where $j=0,1$ is the chosen measurement. The first addend corresponds to the inconclusive result, while the second indicates a successful detection of the state $\ket{\Psi_k(1-j)}$.
The probability of successful discrimination given the measurement is correctly chosen is then
\begin{equation}
\begin{split}
	P_k(0|0) = P_k(1|1) =
	\mbox{Tr}_B\Big[ |\Psi_k(j)\rangle_B{}_B\langle \Psi_k(j)|\\\big( I_k - |\Psi_k(1-j)\rangle_B{}_B\langle \Psi_k(1-j)|\big)\Big]
	= 1- c_k^2
	\end{split}\end{equation}

If the states are modified by Eve, Bob sees correct bits with the probability
\begin{equation}\begin{split}
\mbox{P}_k(j|j)=\mbox{Tr}_B
\Big[
\rho_k(j)\big(I_k-|\Psi_k(1-j)\rangle_B{}_B\langle \Psi_k(1-j)|\big)
\Big]\\
=|A_k|^2\big(1-c_k^2\big),
\end{split}\end{equation}
and error bits with the probability
\begin{equation}\begin{split}
\mbox{P}_k(1-j|j)=\mbox{Tr}_B
\Big[
\rho_k(j)\big(I_k-|\Psi_k(j)\rangle_B{}_B\langle \Psi_k(j)|\big)
\Big]\\
=|B_k|^2\big(1-c_k^2\big).
\end{split}\end{equation}
The observed QBER is then
\begin{equation}
Q_k = \frac{|B_k|^2}{|A_k|^2 + |B_k|^2}.
\end{equation}

Without loss of generality we may assume that $A_k$ and $\braket{\Psi_k(0)|\Psi_k(1)}$ are real. Then, from the normalization of $|\Phi_k(j)\rangle$ it is easy to show that both $B_k$ and $\langle E_k(0) |E_k(1)\rangle$ are also real.
As follows from the unitarity condition of $U_{BE}$,
\begin{equation}
c_k\braket{\Phi_k(0)|\Phi_k(0)} = \braket{\Phi_k(0)|\Phi_k(1)}.
\label{eq_unitarity}
\end{equation}
Substituting~(\ref{eq_unitary_op}) into~(\ref{eq_unitarity}), we obtain
\begin{equation}
c_k\big(1-c_k^\mathrm{E}\big)\big(A^2+B^2\big) = 2(1-c_k^2c_k^\mathrm{E})AB,
\end{equation}
which in turn can be directly tied to the induced QBER level using
\begin{equation}
\frac{2AB}{A^2+B^2} = \sqrt{1-(1-2Q_k)^2 }.
\end{equation}

The desired bound on Eve's states separability given the observed QBER is then
\begin{equation}
c_k^\mathrm{E}(Q_k) =
\frac{c_k - \sqrt{1-(1-2Q_k)^2 }    }
{c_k\left[ 1 -  c_k\sqrt{1-(1-2Q_k)^2   }  \right]  }.
\end{equation}

Having found the maximal separation of the ancilla states for a certain level  of the induced QBER, we can calculate the corresponding bound on the available secret information. After conclusive Bob's measurements, Eve has the following reduced density matrices:
\begin{equation}
\begin{split}
\rho_k^\mathrm{E}(0) = \frac{A^2\ket{E_k(0)}\bra{E_k(0)} + B^2\ket{E_k(1)}\bra{E_k(1)}}{A^2+B^2}\\
\rho_k^\mathrm{E}(1) = \frac{A^2\ket{E_k(1)}\bra{E_k(1)} + B^2\ket{E_k(0)}\bra{E_k(0)}}{A^2+B^2}
\end{split}
\end{equation}
The conditional entropy between Alice and Eve is bounded by the Holevo quantity
$\chi_k(Q_k)$~\cite{H73,H98}:
\begin{equation}
H_k(X|E)\ge 1-\chi_k(Q_k),
\end{equation}
where
\begin{equation}
\chi_k(Q_k)=
H\left( \frac{\rho_k^\mathrm{E}(0) + \rho_k^\mathrm{E}(1)}{2} \right)
-\frac{  H\big(\rho_k^\mathrm{E}(0)\big) + H\big(\rho_k^\mathrm{E}(1)\big) }{2},
\end{equation}

It can be calculated using the following equations:
\begin{equation}
H\left( \frac{\rho_k^\mathrm{E}(0) + \rho_k^\mathrm{E}(1)}{2} \right)
= h\big(\Lambda_k(Q_k)\big),
\end{equation}
where the eigenvalue
$\Lambda^{\pm}_k(Q_k)=
\left( 1   \pm c_k^\mathrm{E}(Q_k)  \right)/2$ and $h(x)=-x\log(x)- (1-x)\log(1-x)$.

Similarly,
\begin{equation}
H\big(\rho_k^\mathrm{E}(0)\big) =  H\big(\rho_k^\mathrm{E}(1)\big)  = h(\lambda_k)
\end{equation}
with the eigenvalue
\begin{equation}
\lambda^{\pm}_k(Q_k)=
\frac{1}{2}
\left(1   \pm  \sqrt{1- 4Q_k \big(1-c_k^\mathrm{E}(Q_k)^2\big)( 1 -  Q_k)    }\right).
\end{equation}

Thus, we have found the expression, that gives the lower bound on the conditional entropy $H_k(X|E)$ for a certain level of the induced QBER. To calculate the secret key rate in realistic conditions we take into account that the observed bit error rates are in general larger than their theoretical bounds, while the detection probabilities are lower. We use the barred versions of corresponding variables to show the observed quantities.

The total probability of detection for the $k$-photon component is
\begin{equation}
\overline{P}_k=\frac12\left(\overline{P}_k(0|0)+\overline{P}_k(1|0)+\overline{P}_k(1|1)+\overline{P}_k(0|1)\right),
\end{equation}
while the corresponding QBER is
\begin{equation}
\overline{Q}_k = \frac{\overline{P}_k(1|0) + \overline{P}_k(0|1)}{2\overline{P}_k}.
\end{equation}
The overall probability of detection is
\begin{equation}
\overline{P} = \sum_{k=0}^{\infty} e^{-2\mu} \frac{(2\mu)^k}{k!} \overline{P}_k,
\end{equation}
and the asymptotic secret key rate equals at least
\begin{equation}
R(\mu)=
\sum_{k=1}^{\infty} e^{-2\mu} \frac{(2\mu)^k}{k!} \overline{P}_k
\big[1-\chi_k(\overline{Q}_k)\big]
-\overline{P}\: h(\overline{Q}),\label{eq_secret_key_rate}
\end{equation}
where $\overline{Q}$ is the total observed QBER. The information $\overline{P}\: h(\overline{Q})$ is the minimal information leakage required for classical error-correction in the Shannon limit.

\section{Decoy state protocol for the practical bound on the secret key rate}

As Alice and Bob do not know the values of $\overline{P}_k$ and $\overline{Q}_k$ for each of the $k$-photon components, practical systems need to comply with a more conservative approach based on measurable parameters only. The decoy state approach~\cite{MQZ05} allows one to find corresponding bounds for the single-photon component, which is enough to guarantee a delivery of secret information.

According to the decoy state strategy, besides the standard pulses with the mean photon number of $\mu$, weaker ``decoy states'' with the corresponding numbers of $\nu_1$ and $\nu_2$ ($0\approx \nu_2 < \nu_1 < \mu$) are randomly chosen. Due to Eve's inability to tell whether a particular pulse was sent as a decoy or as a regular one, a straightforward PNS attack would result in a measurable deviation of the statistics. If we reasonably assume that a $k$-photon component has the same behavior regardless of the original amplitude of the coherent state, we can find direct relations between its detection parameters and some measurable quantities. In particular, it is shown that the minimal probability of detection for a single photon component is given by~\cite{MQZ05}
\begin{equation}\begin{split}
\overline{P}_1\ge
\frac{1/2}{\nu_1-\nu_2 -\frac{\nu_1^2-\nu_2^2}{\mu}}
\Bigg[
\overline{P}(\nu_1) e^{2\nu_1} -  \overline{P}(\nu_2) e^{2\nu_2}
-\\
\frac{\nu_1^2-\nu_2^2}{\mu^2}  \left( \overline{P}(\mu) e^{2\mu} - P_0  \right)
\Bigg],\end{split}
\end{equation}
where
\begin{equation}
\overline{P}_0\ge
\max
\left\{
\frac{ \nu_1 e^{2\nu_2} \overline{P}(\nu_2)  -  \nu_2 e^{2\nu_1}\overline{P}(\nu_1)  }
{\nu_1-\nu_2},0 \right\}.
\end{equation}

Similarly, the upper bound on the single photon component QBER is
\begin{equation}
\overline{Q}_1\le
\frac{  e^{2\nu_1}\overline{P}(\nu_1)\overline{Q}(\nu_1) -   e^{2\nu_2}\overline{P}(\nu_2)\overline{Q}(\nu_2)  }
{ 2(\nu_1 - \nu_2)\overline{P}_1   }.
\end{equation}
In principle, it is possible to find  similar bounds for $k\ge2$, but it would require more than three different amplitudes of the source states~\cite{LMC05}. Not only does it make the whole system less practical, but also the contribution to the overall secret key rate severely decreases with $k$. Therefore, most practical systems rely upon the single photon component only, and effectively, truncate the sum~(\ref{eq_secret_key_rate}) to just only the first term with $k=1$:
\begin{equation}
R'(\mu)=
 2\mu e^{-2\mu} \overline{P}_1
\big[1-\chi_1(\overline{Q}_1)\big]
-\overline{P}(\mu)\: h\big(\overline{Q}(\mu)\big).\label{eq_final_key}
\end{equation}
As one can see, this expression is based on measurable parameters only and, thus, can be used in realistic systems for estimation of the secret key rate. 

\section{Estimation of secret key rates in realistic systems}
It is clear that without Eve's actions the system performance is defined by the linear optical loss in the line and the detector efficiency $\eta$. If the bases of Alice and Bob match, the probability of detection for a single photon component is
$\overline{P}_1 = p_d+ \eta(1-c_1^2) T(L)/2 = p_d + \eta\sin^2(\Delta\varphi/2)T(L)/2$, where $p_d$ is the detector dark count probability per single detection window, $T(L)$ is the overall optical system transmission coefficient, and the factor of $1/2$ is the probability of having a correct measurement choice. Similarly,
\begin{equation}
\overline{P}_k = p_d+\frac12 \bigg(1-\Big[1-\eta\sin^2\left(\Delta\varphi/2\right)T(L)\Big]^k\bigg).
\end{equation}

From the above results it is easy to show that the measured overall detection probability for signal pulses is
\begin{equation}
\overline{P}(\mu)= p_d+\frac12\Big(1-\exp\big[-2\mu\eta\sin^2 (\Delta\varphi/2)T(L)\big]\Big),
\end{equation}
while the corresponding QBER is
\begin{equation}
\overline{Q}(\mu) = \frac{p_d}{2\overline{P}(\mu)}
\end{equation}
The same expressions may be written for decoy states by replacing $\mu$ with $\nu_1$ and $\nu_2$.

This is now enough to estimate the performance of a realistic system. Using the above values for $\overline{P}(x)$ and $\overline{Q}(x)$ ($x={\mu, \nu_1, \nu_2}$), we calculate the ``observed'' bounds on $\overline{P}_1$ and $\overline{Q}_1$, and substitute them into~(\ref{eq_final_key}) to find the asymptotic secret key rate. As an example, we use the following typical values of the parameters: $\eta = 0.1$, $T(L)= 10^{-aL/10}$, where $a=0.2$~dB/km, $p_d = 10^{-6}$. To maximize the efficiency of the parameter estimation for the single photon component, we use the following mean number of photons for decoy states: $2\nu_1 = 0.05$ and $2\nu_2 = 10^{-3}$. Signal states are assumed to have the following mean photon numbers: $2\mu = \{0.1; 0.25; 0.5; 0.9\}$.

To better visualize the results we use the following performance figures: the first is a secret key rate per sifted key bit, defined as $R'(\mu)/\overline{P}(\mu)$. It is of great practical importance as it defines the minimal amount of key material compression required for achieving the goal of key distribution. The second figure is the normalized secret key rate $R'(\mu)/[\eta T(L)]$, which tells how many secret bits we get compared to an ideal single photon based communication line. It allows one to compare secret key generation rates in different configurations.

\begin{figure*}[tb]
	\begin{center}
		\includegraphics[width=0.85\textwidth]{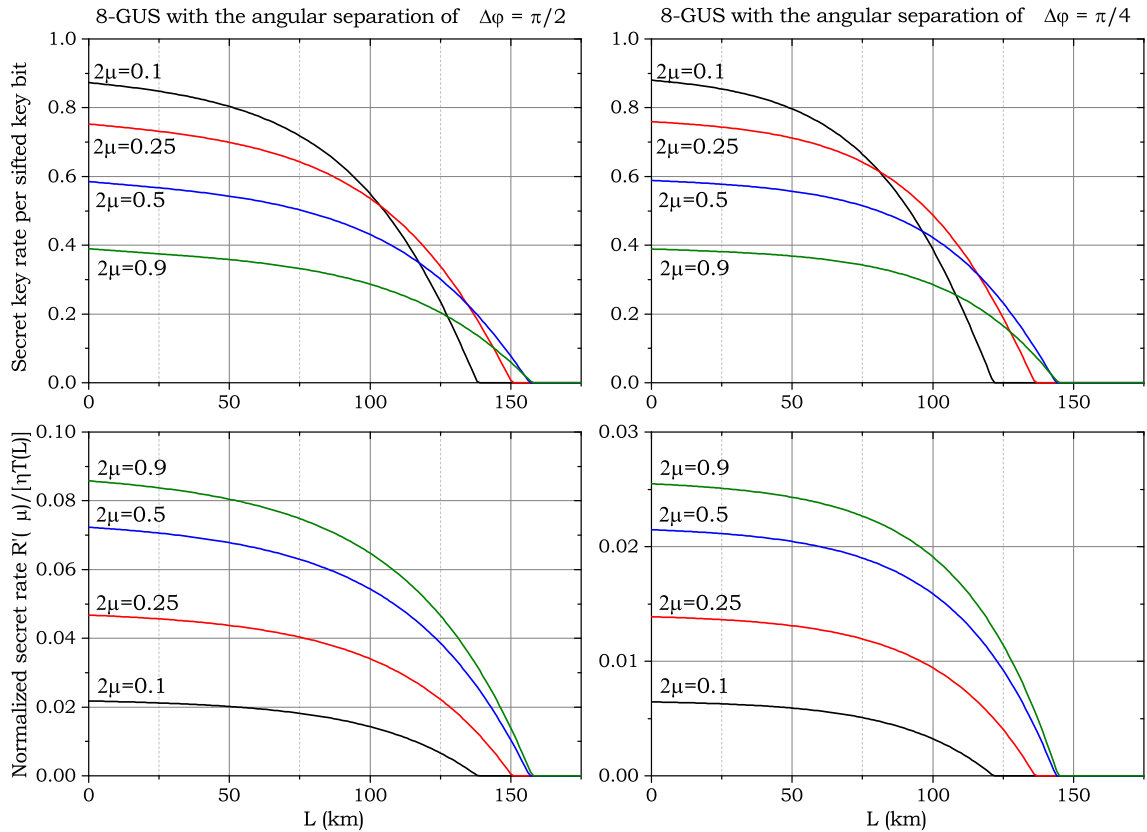}
	\end{center}
	\caption{
		Calculated secret key rate per sifted key bit (top) and the normalized secret key rate (bottom) for the 8-GUS protocol as a function of the channel length $L$. Two columns correspond to different values of $\Delta\varphi$, and curves in each plot vary by the mean photon number $\mu$ for signal states.
		\label{fig_GUS_performance}
	}
\end{figure*}

Calculation results are shown in Fig.~\ref{fig_GUS_performance}. As follows from the plots, the smaller $\mu$ the higher is the secret key rate per sifted key bit. At $\mu=0.1$ almost all the information in the sifted key is secret. It is the consequence of the fact that at such $\mu$ almost all detected states were emitted with just one photon. In contrast, higher $\mu$ leads to the decrease of the single photon component.
The overall key rate normalized to the system efficiency $\eta T(L)$ is higher for larger $\mu$ as more photons reach the detector. In fact the value of $2\mu = 0.9$ is nearly optimal for a channel length of 100~km in terms of the generated amount of secret information.

\section{Discussion}

Construction of a secure and at the same time practical QKD protocol is a hard problem. So far the most popular approach has been the use of the decoy state BB84 protocol. Its construction is directly focused on the protection from the PNS attack, which is also assumed in its standard security analysis. However, from a practical point of view, we must admit that realization of the PNS attack is very far beyond current technological capabilities: it requires both non-demolition photon number measurement and pretty good quantum memory. In contrast, the USD attack may be implemented on-the-fly with quite complex, but available technology. It does not require lossless channels or quantum memory. The measured states may be retransmitted at arbitrary power to compensate for any extra loss between Eve and Bob's detectors. Therefore, it is wise to consider the USD attack to be of more potential danger to practical systems.

At the same time, the USD attack may be successfully used to break the plain WCP-based BB84. 
The decoy state approach is supposed to protect from this attack as well, however specific bounds for USD or its combination with other attacks under this strategy are, to the best of our knowledge, unknown. Moreover, no experimental realization is perfect and any compromising of the decoy state generation hardware directly renders the system insecure even if the adversary is limited by the available technologies, i.e. cannot perform PNS.

In this paper we presented a simple solution implemented at the protocol level, which can be seamlessly combined with the conventional decoy state technique. The provided protocol is not only robust in terms of the negligible probability of the USD, it also extends the ideas coined in~\cite{SAR04} that allow to severely limit Eve's information after a successful PNS attack: logical organization of quantum states into non-orthogonal pairs disallows their deterministic measurement.

It is worth mentioning that the increased number of the proposed logical bases leads to more information loss during key sifting. The same way that the key generation rate in SARG04 is less than in BB84, our protocol is twice less efficient than SARG04 if $\Delta \varphi = \pi/2$ and even worse if it is $\pi/4$. However, the main argument for a QKD system design is its security and not the key generation rate. Modern symmetric ciphers offer unprecedented security level and are widely considered to be as safe even with a quantum computer equipped adversary~\cite{CJL16,MVZ18}. So there is no real need to generate gigabytes of one-time-pad key information: it is more than enough to have symmetric key change once in several seconds. Thus, this minor drawback cannot be significant in real applications. All other parameters, such as the maximal transmission distance, are quite similar to those of conventional protocols.

In conclusion, we constructed a QKD protocol based on $N$ geometrically uniform states. It provides inherent protection from the USD attack by the choice of large enough $N$. Non-orthogonality of logical states in each basis limits the ability of an adversary to perform the PNS attack, which in addition is suppressed by the overlayed decoy state construction. We provide an extensive security analysis of the proposed protocol by considering a general symmetric unitary attack, and give succinct relations for the asymptotic secret key rate in the case of $N=8$. The protocol assumes its straightforward experimental realization by a slight modification of the conventional hardware.

\section*{Acknowledgments}
This work is supported by the Russian Ministry of Education and Science grant No. 03.625.31.0254.

\end{document}